\documentclass[
    aps,
    pra,
   preprint,
    superscriptaddress,
    amsmath,
    amssymb
]{revtex4-2}
\usepackage{graphicx}
\usepackage{dcolumn}
\usepackage{bm}
\usepackage{amsmath}
\usepackage{amssymb}
\usepackage{epstopdf}
\usepackage{xcolor}
\usepackage{comment}
\usepackage[normalem]{ulem}

\graphicspath{{./images/}}
\usepackage{tabularx}
\usepackage{array}
\begin{document}
\newcommand{\js}[1]{{\color{red}#1}}
\newcommand{\kB}{k_{\text{B}}}
\newcommand{\Tc}{T_{\text{c}}}
\newcommand{\nc}{n_{\text{c}}}
\newcommand{\pc}{p_{\text{c}}}


\title{Universal critical behavior in ideal Bose--Einstein condensation}

\author{Arturo Camacho-Guardian}
\email{arturo@fisica.unam.mx}
\affiliation{Instituto de F\'isica, Universidad Nacional
Aut\'onoma de M\'exico, Ciudad de M\'exico, Mexico}

\author{Leon Kleebank}
\author{Frank Vewinger}
\author{Martin Weitz}
\affiliation{Institut f\"ur Angewandte Physik,
Universit\"at Bonn, Wegelerstrasse 8,
53115 Bonn, Germany}

\author{Julian Schmitt}
\affiliation{Kirchhoff-Institut f\"ur Physik,
Universit\"at Heidelberg, Im Neuenheimer Feld 225a,
69120 Heidelberg, Germany}

\author{Rosario Paredes}
\author{Victor Romero-Roch\'{\i}n}
\affiliation{Instituto de F\'isica, Universidad Nacional
Aut\'onoma de M\'exico, Ciudad de M\'exico, Mexico}
\email{arturo@fisica.unam.mx}\

\begin{abstract}
The experimental realization of photonic Bose-Einstein condensates
has renewed the interest in the critical behavior of the ideal Bose gas,
a paradigmatic system generally considered to be well
understood. Here we establish a universal thermodynamic classification of ideal Bose-Einstein condensation in terms of experimentally measurable critical observables. We identify three universality classes with distinct critical exponents and singular thermodynamic behavior, arising from the low-energy scaling of the single-particle density of
states. Depending on the low-energy scaling exponent of the density of states, which is controlled by dimensionality and confinement, the transition displays either the usual algebraic divergences of thermodynamic susceptibilities, divergent behavior with marginal logarithmic corrections, or a more subtle form of criticality, where only the correlation length diverges. Our work provides a unified framework for equilibrium criticality in noninteracting bosonic systems. This classification applies broadly to atomic, photonic, polaritonic, and magnonic condensates, where dimensionality, confinement, and spectral engineering can strongly reshape the density of states. We connect our thermodynamic classification with previously identified universality classes based on condensate fluctuations and order parameters.
\end{abstract}
\maketitle

\section{ Introduction} 
The experimental realization of Bose–Einstein condensation (BEC) in photon gases has opened new opportunities to investigate critical phenomena in nearly ideal quantum systems~\cite{Klaers2010,Marelic:2015,Greveling:2018}. In dye-filled optical microcavities, photons thermalize through repeated absorption and re-emission by dye molecules, while their effective mass, dimensionality, and confinement can be engineered experimentally. These features make photonic condensates particularly attractive for exploring the critical behavior of essentially noninteracting Bose gases. Recent advances have allowed the study of condensation beyond signatures of macroscopic occupation of the ground state, providing access to the equation of state, compressibility, and spatial correlations~\cite{Damm:2016,Busley2022,Kleebank:2025}. Together with the possibility of tailoring the trapping potential~\cite{Busley2022,Umesh2024}, these developments raise a fundamental question: How does the geometry of confinement determine the universal critical behavior of an ideal Bose gas?

Critical exponents characterize the singular behavior of thermodynamic susceptibilities and correlation functions near continuous phase transitions~\cite{Fisher1974}. In interacting systems, critical behavior is governed by collective fluctuations and often requires theoretical descriptions beyond mean-field approximations~\cite{Wilson1974,Ma1976,Amit1984}. For instance, interacting Bose gases exhibit critical behavior associated with the XY universality class~\cite{Lipa,Burovski}, while two-dimensional systems can undergo Berezinskii–Kosterlitz–Thouless transitions~\cite{Kosterlitz1974}. In contrast, the critical properties of an ideal Bose gas arise solely from quantum statistics and the structure of its single-particle spectrum. Although ideal BEC is one of the paradigmatic examples of a phase transition arising from quantum statistics, its critical behavior can be remarkably sensitive to dimensionality and confinement.

Indeed, BEC has been realized in a broad range of physical systems, including ultracold atomic gases, exciton-polaritons, magnons, and photons~\cite{Cornell1995,Ketterle1995,Deng2010,Demokritov2006,Klaers2010,Schofield:2024,Pieczarka:2024}. However, these experimental realizations typically differ substantially from the textbook picture of a homogeneous, noninteracting three-dimensional gas. Atomic condensates generally involve interactions and external trapping potentials, whereas polaritonic and magnonic condensates are often driven-dissipative systems. Photon gases in dye-filled microcavities offer a particularly appealing setting for investigating equilibrium Bose statistics, as coupling to the dye bath enables thermalization while effective interactions remain weak and confinement can be controlled. Crucially, changing dimensionality or the trapping potential modifies the low-energy single-particle density of states (DOS), suggesting a direct connection between spectral engineering and critical thermodynamic behavior.

The role of dimensionality and confinement in determining the existence of ideal BEC has long been recognized~\cite{Bagnato1987,LandauLifshitz}. Previous studies of criticality and universality in ideal Bose gases have identified three distinct classes based on the statistical properties of the excited-state particle-number fluctuations~\cite{Tarasov,Tarasov2015}, while other approaches have investigated critical behavior through phenomenological order parameters~\cite{Gunton}. Yet, a systematic characterization of these classes in terms of thermodynamic critical exponents, scaling relations, and directly measurable observables, such as thermodynamic susceptibilities and spatial correlations, remains to be established. Such a framework is particularly timely in light of recent experimental advances in photonic Bose gases.

In this work, we establish three distinct classes of critical behavior for ideal Bose–Einstein condensation, determined entirely by the low-energy scaling of the single-particle DOS, $\rho(\epsilon)\sim\epsilon^\sigma$. For $0<\sigma<1$, thermodynamic susceptibilities exhibit algebraic divergences; at the marginal point $\sigma=1$, logarithmic corrections emerge; whereas for $\sigma>1$, the compressibility remains finite even though the correlation length diverges. We derive the corresponding critical exponents and scaling relations, and connect this thermodynamic classification with previously identified regimes of particle-number fluctuations. Our results demonstrate how dimensionality and confinement alone can determine the universal properties of the BEC transition, providing a framework directly relevant to current experiments on photonic quantum gases.

\begin{figure}[t]
    \centering
    \includegraphics[width=.75\linewidth]{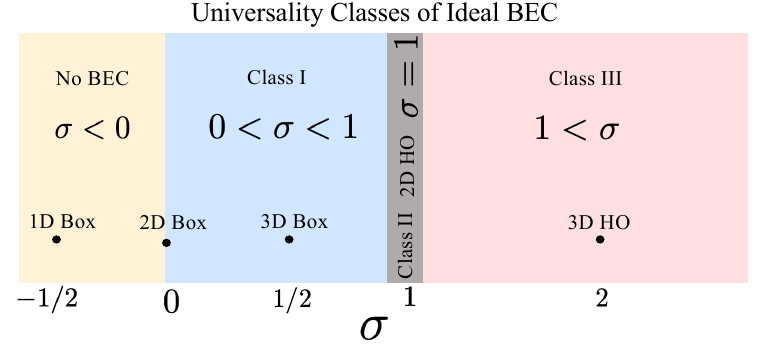}
    \caption{Classification of universal BEC critical behavior as a function of the low-energy scaling behavior of the density of states $\rho(\epsilon) \sim \epsilon^\sigma$, quantified by $\sigma$. Black dots indicate example systems. For $\sigma\le 0$ (\textit{e.g.}, Bose gas in 1D or 2D box traps), BEC does not occur at finite temperature (yellow box). For $0<\sigma<1$ (Class~I, \textit{e.g.}, 3D box), the thermodynamic susceptibilities diverge (blue). The marginal case $\sigma = 1$ [Class~II, \textit{e.g.}, 2D harmonic oscillator (HO) trap] features logarithmic corrections to the divergent scaling (gray). For $\sigma > 1$ (Class~III, \textit{e.g.} realized in a 3D harmonic trap), only the correlation length diverges with the exponent $\nu_T = 1/2$, while the isothermal compressibility remains finite.}
    \label{fig:sigma_phase_diagram}
\end{figure}

\vspace{0.2cm}
\section{ Scaling laws of ideal BEC} The thermodynamic properties of the ideal Bose gas at temperature $T$ are governed by the grand potential~\cite{LandauLifshitz},
\begin{equation}
\Omega = \kB T \int_0^\infty \rho(\epsilon)\,
\ln\!\left(1-e^{- \epsilon/\kB T + \tilde\mu}\right)
d\epsilon,
\label{Omega}
\end{equation}
with  $\tilde\mu =\mu/\kB T$, where $\mu \le 0$ is the chemical potential. The single-particle DOS $\rho(\epsilon)$ is fixed in the thermodynamic limit and encodes information about the confining potential $V(\mathbf r)$, \textit{e.g.} harmonic, box, power-law, or soft-wall Pöschl-Teller potentials~\cite{Poschl1933,Nieto1978,Kleebank:2025}. It scales linearly with an extensive generalized volume ${\cal V}$~\cite{RomeroRochin2005,Sandoval2008}, may depend on intensive potential parameters, and is strongly influenced by the dimensionality $d$; for a detailed discussion, see the Appendix.

The onset of BEC at temperature $T$ is controlled by the particle density $n=N/{\cal V}$,
\begin{eqnarray}
n &=& -\frac{1}{{\cal V}} \left(\frac{\partial \Omega}{\partial\mu}\right)_{T,\mathcal V} = \frac{1}{{\cal V}} \int_0^\infty 
\frac{\rho(\epsilon)}{e^{ \epsilon/\kB T - \tilde\mu}-1} d\epsilon  \>.
\label{nT}
\end{eqnarray}
At $\tilde\mu=0$, the gas exhibits BEC if $n$ remains finite, and this occurs only if the low-energy DOS vanishes sufficiently rapidly with energy $\epsilon$. Writing
$\rho(\epsilon) \sim \epsilon^\sigma$, the integral diverges for $\sigma \le 0$ and BEC is impossible at finite $T$ and $n$; for example, a homogeneous Bose gas in a 2D box trap has $\sigma=0$ and does not condense, see Fig.~\ref{fig:sigma_phase_diagram}. The scaling exponent $\sigma > 0$ alone dictates whether ideal-gas condensation can occur. The role of the low-energy DOS in determining the existence of Bose--Einstein condensation has long been recognized. Refs.~\cite{Bagnato1987,Bagnato1991} established the conditions for condensation in arbitrary trapping potentials, while Ref.~\cite{Tarasov} investigated universal statistical and thermodynamic properties in arbitrary traps. In interacting Bose systems, criticality is conventionally described in terms of an order parameter arising from spontaneous U(1) symmetry breaking. Motivated by this framework, Ref.~\cite{Gunton} analyzed the critical behavior of the ideal Bose gas with translational symmetry by introducing a phenomenological Bose order parameter. For the ideal Bose gas, however, such a field is not required within equilibrium thermodynamics. Here, we instead formulate the critical behavior entirely in terms of measurable thermodynamic observables, without introducing an order parameter.  

At $\epsilon\to 0$, the DOS can quite generally be written as
\begin{equation}
\rho(\epsilon) = {\cal V} C \epsilon^\sigma + \textrm{higher powers in $\epsilon$} ,
\end{equation}
where $C$ depends on fundamental constants, the mass of the particles, and intensive trap parameters, if any. The exponent $\sigma$ is fixed entirely by the low–energy part of the confining potential. This can be seen by semiclassically evaluating the DOS,
\begin{equation}
\rho(\epsilon)
= \frac{1}{(2\pi\hbar)^d}
\int d^d r\, d^d p\;
\delta\!\left(\epsilon - \frac{p^2}{2m} - V(\mathbf r)\right) \>.
\label{eq:WeylDOS}
\end{equation}
To characterize the corresponding low-energy behavior, we measure energy from the global minimum of the external potential $V(\mathbf r)$, which we set to zero for simplicity. Since only the local form of the potential near its minimum contributes to the DOS as $\epsilon \to 0^+$, we approximate
\begin{equation}
V(\mathbf r) \simeq V_0 r^s \>,
\label{eq:localV}
\end{equation}
where $s$ characterizes the power–law confinement. We assume isotropy; the anisotropic generalization is given in the Appendix. Inserting Eq.~\eqref{eq:localV} into Eq.~\eqref{eq:WeylDOS}, one finds that the spatial region accessible to energy $\epsilon$ scales as $\epsilon^{d/s}$, giving the universal infrared form $\rho(\epsilon \to 0^+) \propto \mathcal V\, \epsilon^\sigma$ with the scaling exponent
\begin{equation}
\sigma = \left(\frac{d}{2} - 1\right) + \frac{d}{s},
\label{eq:scaling_sigma}
\end{equation}
which shows explicitly that $\sigma$ is determined by confinement and dimensionality. Note that $\sigma$ is independent of the details of $V(\mathbf r)$ away from the minimum. Here, the generalized volume associated to the potential $V(\mathbf r)$ in Eq.~\eqref{eq:localV} scales as $\mathcal V \sim V_0^{-d/s}$, see Appendix. Thus, the DOS near threshold is governed solely by the local curvature or flatness of the potential. Equation~\eqref{eq:scaling_sigma} reproduces familiar limits: For a box ($s\to\infty$), one finds $\rho(\epsilon)\propto \epsilon^{d/2-1}$ allowing BEC in $d> 2$, while for harmonic confinement ($s=2$), one obtains $\rho(\epsilon)\propto\epsilon^{d-1}$, allowing BEC in $d\geq 2$~\cite{Bagnato1987,PethickSmith2008,MoralesAmador2024}. Of particular importance is the case $\sigma=1$, which is realized by an ideal photon gas confined in a harmonic potential~\cite{Klaers2010}, or in a box-like potential with softened edges~\cite{Kleebank:2025}.

As the BEC limit is realized by $\tilde\mu\to 0^-$, we can write, see Eq.~(\ref{nT}),
\begin{eqnarray}
n =C_0 T^{\sigma+1} g_{\sigma+1}(\tilde\mu) +\dots,
\label{n}
\end{eqnarray}
where $C_0 = C k_B^{\sigma+1} \Gamma(\sigma+1)$ with Gamma function $\Gamma(x)$. The Bose function
\begin{equation}
g_{\sigma + 1}(\tilde\mu) = \frac{1}{\Gamma(\sigma+1)}
\int_0^\infty
\frac{x^\sigma}{e^{x - \tilde\mu}-1} dx \>
\end{equation}
is non-analytic at $\tilde\mu=0$ and for $\sigma > 0$ and $|\tilde\mu| < 1$ can be expressed by asymptotic expansions~\cite{ZUK-BEC,Robinson1951}:
\begin{widetext}
\begin{subequations}
\begin{eqnarray}
g_{\sigma + 1}(\tilde\mu) &=& \Gamma(-\sigma)(-\tilde\mu)^\sigma + 
\sum_{k=0}^\infty \frac{\zeta(\sigma+1-k)}{k!}\tilde\mu^k 
\>\>\>\>\>\sigma \ne 1,2,3,\dots \label{asin2}\\
g_{\sigma + 1}(\tilde\mu) &=& \frac{\tilde\mu^{\sigma}}{\sigma!}
\left[H_\sigma - \ln(-\tilde\mu) \right] +
\sum_{k=0, k\ne \sigma}^\infty 
\frac{\zeta(\sigma+1-k)}{k!}\tilde\mu^k 
\>\>\>\>\>\sigma = 1,2,3,\dots\>
\label{asin}
\end{eqnarray}
\end{subequations}
\end{widetext}
Here, $H_\sigma$ denotes the harmonic number and $\zeta$ the Riemann zeta function. For $\sigma>0$, one has the critical phase-space density $g_{\sigma+1}(0)=\zeta(\sigma+1)$.

Substituting Eqs.~\eqref{asin2} and \eqref{asin} into the expression Eq.~\eqref{n} for the density $n$ and expanding near criticality ($\tilde\mu\to 0^-$), we obtain three quantitatively different regimes of {criticality},
\begin{equation}
\zeta(\sigma + 1) t \approx 
- \left\{
\begin{array}{ccc}
 \Gamma(-\sigma) (-\tilde\mu)^\sigma + \cdots & (\sigma < 1) \\
 & & \\
 -\tilde\mu \ln(-\tilde\mu) + \cdots & (\sigma = 1) \\
 & & \\
 \zeta(\sigma) \tilde\mu + \cdots & (\sigma > 1)
\end{array}\right.\!,
\label{t-cases}
\end{equation}
Here $t$ denotes the reduced temperature or density, respectively,
\begin{equation}
t = (\sigma+1)\frac{T-\Tc}{\Tc} = \frac{\nc-n}{\nc}.
\end{equation}
relative to the critical point. Fixing either $n$ or $T$ yields finite critical values $\Tc$ or $\nc$, respectively. Specifically, in the critical regime $t\ll 1$. Note that $\zeta(\sigma + 1)=(n/C_0T^{\sigma+1})_c$ is the critical phase-space density. Equation (\ref{t-cases}) allows to solve for the reduced chemical potential in terms of $t$,
\begin{equation}
-\tilde\mu \approx \left\{
\begin{array}{ccc}
 {\cal A}_{\tilde\mu} \, t^{1/\sigma} & (\sigma < 1) \\
 & & \\
 -{\cal B}_{\tilde\mu} \, t (\ln t)^{-1} & (\sigma = 1) \\
 & & \\
 {\cal C}_{\tilde\mu} \, t & (\sigma > 1)
\end{array}\right.\!.
\label{alf-cases}
\end{equation}
As all thermodynamic quantities depend on the chemical potential $\tilde\mu$, critical exponents can be extracted.

We first consider the isothermal compressibility $\kappa_T = n^{-2} (\kB T)^{-1}
({\partial n}/{\partial \tilde\mu})_T$, which follows from Eq.~\eqref{n}. One finds
\begin{equation}
\kappa_T \approx 
\left\{
\begin{array}{ccc}
 {\cal A}_\kappa \, t^{-(1/\sigma - 1)} & (\sigma < 1) \\
 & & \\
 -{\cal B}_\kappa \ln t & (\sigma = 1) \\
 & & \\
 {\cal C}_\kappa + \cdots & (\sigma > 1)
\end{array}\right.\!\>.
\label{kapa}
\end{equation}
Thus, the obtained algebraic scaling form $\kappa_T \sim t^{-\gamma_T}$ implies a critical exponent $\gamma_T = 1/\sigma - 1$ for $\sigma < 1$. At $\sigma = 1$, the compressibility shows a logarithmic divergence. By contrast, for $\sigma>1$, the compressibility does not diverge at all, so that no $\gamma_T$ exponent can be assigned.

Similarly, we analyze the critical scaling of the pressure $p$, deduced from $\Omega=-p{\cal V}$ and Eq.~\eqref{Omega},
\begin{equation}
p  \approx  k_B C_0 T^{\sigma+2}
\left[\zeta(\sigma+2) + \zeta(\sigma+1)\tilde\mu + \cdots\right].
\end{equation}
At $T=\Tc$, using Eq.~(\ref{alf-cases}),
\begin{equation}
p - \pc \approx 
\left\{
\begin{array}{ccc}
 {\cal A}_p (n - \nc)^{1/\sigma} & (\sigma < 1) \\
 & & \\
 {\cal B}_p (n - \nc) \frac{1}{\ln(n-\nc)} & (\sigma = 1) \\
 & & \\
 {\cal C}_p (n - \nc) & (\sigma > 1)
\end{array}\right.\!.
\label{p}
\end{equation}
Thus, $p-\pc\sim (n-\nc)^{\delta_{T}}$ gives a critical exponent $\delta_T = 1/\sigma$ for $\sigma\le1$, with logarithmic corrections at $\sigma=1$, and the mean-field-like value $\delta_T=1$ for $\sigma>1$. Finally, note that the heat capacity $C_{\mathcal V}$ for ideal Bose gases at constant volume $\mathcal V$ does not diverge at $\Tc$ for any confinement. This is because, near BEC, it scales with $g_{2+\sigma}(\tilde \mu) \sim \zeta(2+\sigma)$, which is finite for any value of $\sigma$. One finds $C_{\cal V} \sim \mathcal V T^{\sigma + 1} \sim N$ at $\mu = 0$. Therefore, a critical exponent associated to the heat capacity may not be meaningful. 

\vspace{0.2cm}
\section{Correlations and fluctuations at criticality} A key observable to describe the critical behavior of a quantum gas near a phase transition is the density–density correlation function $G(\textbf{x},\textbf{x}')=\langle n(\textbf{x})n(\textbf{x}')\rangle- \langle n(\textbf{x})\rangle \langle n(\textbf{x}')\rangle$ and its associated correlation length $\xi$. Inhomogeneous systems, \textit{e.g.}, in harmonic trap potentials, break translational invariance, so that $G(\mathbf x,\mathbf x')\neq G(\mathbf x-\mathbf x')$. The correlations then depend separately on the relative and the center–of–mass coordinates, $\mathbf r = \mathbf x - \mathbf x',$ and $\mathbf R = (\mathbf x + \mathbf x')/2$. Within the local density approximation (LDA), the subsystem at $\mathbf R$ is treated as a homogeneous gas with a locally varying chemical potential
\begin{equation}
\mu_{\mathrm{LDA}}(\mathbf R)=\mu - V(\mathbf R)\>.
\label{eq:mu_LDA}
\end{equation}
For a homogeneous ideal Bose gas in $d$ dimensions in the quantum-degenerate regime ($\mu \to 0^-$), the density–density correlation function is(see Appendix),
\begin{equation}
G_{\mathrm{hom}}(r)
\approx \mathcal A_d\left|\>\left(\frac{1}{\xi r}\right)^{\frac{d}{2}-1}
K_{\frac{d}{2}-1}\!\left(\frac{r}{\xi}\right)\right|^{2},
\label{eq:Ghom}
\end{equation}
with a temperature-dependent correlation amplitude $\mathcal A_d$, correlation length $\xi = \sqrt{\hbar^2/2m|\mu|}$,  $K_\ell(z)$ the modified Bessel function of second kind of order $\ell$. The correlator in Eq.~\eqref{eq:Ghom} depends on the relative coordinate $r$ only, indicating that the system is translationally-invariant. Correspondingly, for an inhomogeneous system the LDA yields
\begin{equation}
G_{\mathrm{LDA}}(\mathbf r,\mathbf R)\approx
\mathcal A_d  \left|\left(\frac{1}{\xi(\mathbf R)|\mathbf r|}\right)^{\frac{d}{2}-1}
K_{\frac{d}{2}-1}\!\left(\frac{|\mathbf r|}{\xi(\mathbf R)}\right)\right|^{2} \>,
\label{eq:GLDA}
\end{equation}
with the local correlation length
\begin{equation}
\xi^{-1}(\mathbf R)=\sqrt{\xi^{-2} + \frac{2m V(\mathbf R)}{\hbar^2}}.
\end{equation}
Integrating out the center-of-mass coordinate $\mathbf R$, using Eq.~\eqref{eq:localV}, gives us the isotropic correlation function between two distinct locations in the trap,
\begin{equation}
\mathcal G(r)=\int d^{d}R\,G_{\mathrm{LDA}}(r,\textbf{R}) \>.
\label{eq:G2_def}
\end{equation}

For large but finite values of $\xi$, Eq.~\eqref{eq:G2_def} can be evaluated using the large–argument form, $r/\xi\gg 1$, of the modified Bessel function(see Appendix). One obtains the density-density correlation function
\begin{equation}
\mathcal G(r)\sim \mathcal V\; \frac{\xi^{-(d-3+d/s)}}{r^{\,d-1+d/s}}\,e^{-2r/\xi} \>.
\label{eq:g2_xi_large}
\end{equation}
which indicates that the spatial correlations exhibit a self-similar power-law decay  with an exponent determined by the confining potential as the correlation length $\xi\to\infty$ very close to BEC. The correlation length itself diverges as $\xi\sim t^{-\nu_T}$. Using Eq.~(\ref{alf-cases}),
\begin{equation}
\xi \approx 
\left\{
\begin{array}{ccc}
 {\cal A}_\xi \, t^{-1/2\sigma} & (\sigma < 1) \\
 & & \\
 {\cal B}_\xi \, [t^{-1} \ln t]^{1/2} & (\sigma = 1) \\
 & & \\
 {\cal C}_\xi \, t^{-1/2} & (\sigma > 1)
\end{array}\right.\!.
\label{xi-cases}
\end{equation}
Here, clearly the exponent $\nu_T = 1/2\sigma$ for $\sigma\le1$ is universal for all systems showing the same scaling behavior of the DOS; for example, from Eq.~\eqref{eq:scaling_sigma}, the 1D Bose gas in a V-shaped linear trap ($d=1,s=1$) exhibits the same universality as the 3D box-trapped gas ($d=3,s\rightarrow\infty$), both having $\sigma=1/2$. At $\sigma=1$, the divergence is modified by logarithmic corrections and loses its $\sigma$-dependence; the values are in good agreement with recent experimental results in photon and polariton systems~\cite{Kleebank:2025,Gnusov:2026}. For $\sigma>1$, as seen before for the thermodynamics, the mean-field value $\nu_T=1/2$ is obtained. The result from Eq.~\eqref{xi-cases} highlights that BEC in the ideal gas strictly speaking does not generally correspond to mean-field-like critical behavior, but has a richer, fluctuation-governed criticality that depends sensitively upon the low-energy physics.

\begin{table}[t]
    \centering
    \resizebox{\columnwidth}{!}{%
    \begin{tabular}{c|c|c|ccccc}
    \hline\hline
    Class & Range & BEC  & $\gamma_T$ & $\nu_T$ & $\eta_T$ & $\delta_T$ \\
    \hline
    - (No BEC) & $\sigma\le 0$ & \text{\sffamily X} & - & - & - & - &  \\
    \hline
    I & $0<\sigma<1$ & \checkmark  &
    $1/\sigma-1$ &
    $1/(2\sigma)$ &
    $2\sigma$ &
    $1/\sigma$ \\
    \hline
    II (marginal) & $\sigma=1$ & \checkmark &
    $\sim\!-\ln t$ &
    $\sim\![t^{-1}\ln t]^{1/2}$ &
    $2$ (from $\gamma_T=\nu_T(2-\eta_T)$ with logs) &
    $1$ \\
    \hline
    III & $\sigma>1$ & \checkmark  &
    finite (no divergence) &
    $1/2$ &
    - &
    $1$ \\
    \hline\hline
    \end{tabular}%
    }
    \caption{Overview of the classification for critical behavior of ideal BEC along with the critical exponents based on the reported thermodynamics analysis. The exponents refer to compressibility $\kappa_T\sim t^{-\gamma_T}$, correlation length $\xi\sim t^{-\nu_T}$, density-density correlations ${\cal G}(r)\sim r^{-d+2-\eta_T}$, and pressure $p\sim |n-n_c|^{\delta_T}$.}
    \label{tab:universality_classes}
\end{table}
    
\vspace{0.2cm}
\section{ Scaling relations} To obtain a prediction for the algebraic decay of the spatial correlation function ${\cal G}(r)$ at criticality, $\mu = 0$, and identify a critical exponent $\eta_T$~\cite{Fisher1974}, one can use, first, the limit $r/\xi \ll 1$ of the Bessel function $K_{d/2-1}(r/\xi)$ and, then take the strict limit $\xi \to \infty$. In this case, the local correlation length scales as $\xi^{-1}(R)\!\sim\! {\cal V}^{-s/2d} R^{s/2}$ and the correlation function becomes(see Appendix), 
\begin{equation}
\mathcal G(r)\approx \mathcal V \frac{\mathcal D_d}{r^{d-2+\eta_T}},
\end{equation}
where the critical exponent $\eta_T$ is identified as, 
\begin{equation}
\eta_T = d - 2 + \frac{2d}{s} = 2\sigma.
\label{eq:eta_final}
\end{equation}
The isothermal compressibility and the correlation function are tied by the fluctuation–response relation, 
\begin{align}
\kappa_{T} = \frac{1}{n^2k_BT} \int d^{d}r\, \mathcal G(r) ,
\label{eq:kappa_def}
\end{align}
which when used in conjunction with the asymptotic behavior of $\mathcal G(r)$ at large $r$, Eq.~\eqref{eq:g2_xi_large}, gives the scaling $\kappa_T \sim \xi^{2-\eta_T}$. This yields the Fisher scaling relation $\gamma_T=\nu_T(2-\eta_T)$, which is satisfied for $\sigma \le 1$, as can be verified from the values of the exponents just found. For $\sigma > 1$ the compressibility and the integral of the correlation function no longer diverge and, therefore, the Fisher equality does not apply. Remarkably, the Fisher relation remains valid for trapped ideal gases, even though $\eta_T$ originates from spatial inhomogeneity rather than from interactions. Note that $\eta_T = 2\sigma$ characterizes the density–density response function and should not be confused with the anomalous dimension $\eta$ of the Bose field~\cite{Gunton,Tarasov}, which in the ideal homogeneous gas vanishes, $\eta = 0$, as we have verified in the  Appendix.

Our final result for the critical exponents that classify the ideal BEC transition is summarized in Table~\ref{tab:universality_classes}. In contrast to phenomenological order-parameter approaches for translationally invariant ideal Bose gases~\cite{Gunton}, our classification is derived entirely within equilibrium thermodynamics.  Beyond translationally invariant systems, the two approaches generally lead to different critical exponents. The complete set of critical exponents follows directly from measurable thermodynamic observables, without introducing a phenomenological Bose field.  It is of interest to point out that, for a homogeneous gas in a box potential, where $s \to \infty$, the scaling is $\sigma = d/2-1$. Conversely, for an inhomogeneous confining potential with arbitrary $\sigma > 0$, one can readily define an effective dimension $d_{\text{eff}} = 2 (\sigma + 1)$. This indicates that the algebraic critical behavior $0 < \sigma < 1$ corresponds to $2 < d_{\text{eff}} < 4$, then $d_{\text{eff}} = 4$ is the upper border dimension with logarithmic corrections and $d_{\text{ eff}} > 4$ is of "mean-field" behavior, in agreement with general ideas of critical phenomena. Interestingly, we find 
\begin{equation}
\delta_T = \frac{d_{\text{eff}} + 2 -\eta_T}{d_{\text{eff}} - 2 + \eta_T}
\label{BKT}
\end{equation}
which is universal and found for all critical phenomena of uniform systems, including the Berezinskii–Kosterlitz–Thouless transition~\cite{Kosterlitz1974}. 

Our classification is closely related to the fluctuation-based universality classes identified in Refs.~\cite{Tarasov,Tarasov2015}. The connection between the two classifications follows directly from the fluctuation-response relation $\langle(\Delta N)^2\rangle = k_B T\,n^2 V\,\kappa_T$.
For an ideal Bose gas, the infrared contribution to the number fluctuations is controlled by
\begin{equation}
    \langle(\Delta N)^2\rangle_{\rm IR}     \sim \int_0 d\epsilon\, \rho(\epsilon)\,n_B(\epsilon)    \bigl[1+n_B(\epsilon)\bigr].
\end{equation}
Since $\rho(\epsilon)\sim\epsilon^\sigma$ and $n_B(\epsilon)\sim k_BT/\epsilon$ at low energies, the integrand scales as $\epsilon^{\sigma-2}$. The infrared integral is therefore convergent for $\sigma>1$, logarithmically divergent at $\sigma=1$, and algebraically divergent for $0<\sigma<1$. Thus, the same infrared singularity that determines whether particle-number fluctuations are Gaussian or anomalous also controls whether the thermodynamic compressibility remains finite or diverges. Consequently, the Gaussian-anomalous boundary identified in Refs.~\cite{Tarasov,Tarasov2015} coincides with the marginal point $\sigma=1$ separating finite and divergent thermodynamic susceptibility in our formulation. That is,  our classification in terms of critical exponents and correlation functions is consistent with the universality classes previously identified. Our approach, however, provides a thermodynamic perspective through the direct calculation of critical exponents and the establishment of scaling relations, including the Fisher relation and Eq.~\eqref{BKT}.
This unified framework connects the critical behavior of ideal Bose gases with experimentally accessible thermodynamic susceptibilities and spatial correlations, making it particularly relevant in light of recent experiments with photon gases in optical microcavities.

Two-dimensional photon gases therefore allow one to span the different universality classes by engineering the confining potential: a linear
potential realizes Class III, harmonic and P\"oschl--Teller confinement probe the marginal Class II, while Class I can be accessed using an $r^4$ trap, which can likewise be imprinted onto the cavity mirrors.
Remarkably, recent experiments on a two-dimensional photon gas confined by a P\"oschl--Teller potential have measured a critical correlation-length exponent consistent with the value predicted for our marginal Class II
universality class~\cite{Kleebank:2025}. This provides experimental evidence that ideal Bose--Einstein condensation exhibits a distinct
critical scaling behavior, different from the BKT and three-dimensional
XY universality classes characteristic of interacting Bose systems.
Photon Bose gases therefore provide a particularly suitable platform
for testing the connection between confinement, the low-energy density
of states, and thermodynamic critical scaling.

\vspace{0.2cm}
\section{ Conclusions}  Our work establishes a unifying classification of ideal Bose gases near the  continuous (\textit{i.e.}, second-order) BEC phase transition.
The classification criterion is based solely on the low-energy scaling of the DOS, $\rho(\epsilon)\sim\epsilon^\sigma$. Our analysis demonstrates that the exponent $\sigma$ not only determines whether BEC is thermodynamically allowed but also, if the case, uniquely predicts the critical exponents. Three classes emerge: (i) $\sigma < 1$, a fluctuation-governed regime where the compressibility diverges and all standard critical exponents are well defined; (ii) $\sigma = 1$, a marginal regime where logarithmic corrections appear; and (iii) $\sigma > 1$, a regime with critical exponents resembling that of the mean-field prediction where only the correlation length diverges with the exponent $\nu_T = 1/2$, $\delta_T = 1$, while other familiar exponents such as $\gamma_T$ and $\eta_T$ cease to be meaningful.

In a broader perspective, our findings reveal that the non-analytic behavior of thermodynamic quantities at the BEC point originates from geometry and spectral shaping alone. Besides quantum fluids of light, where recent experiments have obtained results consistent with the presented theory~\cite{Kleebank:2025,Gnusov:2026}, the investigation here exposed is particularly relevant for condensed matter platforms including the emerging designs of band structures, flat bands in layered systems, synthetic dimensions or even fractal lattices where the DOS near the ground state can be tuned deliberately~\cite{DanieliAndreanovLeykamFlach2024,Luengo2024,Dutkiewicz2026}. Extensions of this approach to both weakly interacting as well as nonequilibrium systems may further illuminate how universality and critical behavior evolves from the ideal-gas paradigm to the full complexity of quantum critical phenomena in bosonic matter\cite{Sieberer:2025}, encompassing physical realizations including ultracold atoms in uniform traps~\cite{Navon:2021}, exciton-polaritons in engineered semiconductor microcavities~\cite{Bloch:2022}, or photons in dye-filled cavities~\cite{Klaers2010}. Our classification therefore provides a common language to compare BEC transitions across platforms where the single-particle spectrum can be highly nontrivial.

{\it Acknowledgments.} We thank A. Ran\c{c}on for fruitful discussions. A.C.-G. acknowledges financial support from UNAM DGAPA PAPIIT Grant No. IA101325 and Project SECIHTI No. CBF2023-2024-1765, M.W., F.V. and J.S. from DFG within SFB/TR 185 (277625399) and Cluster of Excellence ML4Q (EXC 2004/1, 390534769), and J.S. from the EU (ERC, TopoGrand, 101040409) and the DFG within Cluster of Excellence STRUCTURES (EXC 2181, 390900948). V.R.-R. and R.P. acknowledge funding from UNAM DGAPA PAPIIT Grant IN10772, and A.C.-G. and R.P. from PIIF25.

\appendix
\begin{widetext}
\section{ Thermodynamic Limit and Density of States at low lying energies for arbitrary confining potentials}

In this section we derive the low lying energy expression of the density of states (DOS) for a generic potential, using the known semiclassical expression and, in the following section, we provide examples with standard confinement geometries. With this procedure, we can (a) identify the generalized volume $\mathcal V$, corresponding to the particular external confining potential, and that defines the thermodynamic limit and (b) extract the low-energy power-law exponent $\sigma$. We show that the low-energy DOS takes the form, 
\begin{equation}
    \rho(\epsilon) \xrightarrow[\epsilon \to 0^+]{} {\cal V}\, C\, \epsilon^{\sigma},
\end{equation}
with $C$ a constant. As shown in the main text, this DOS expression completely determines both the existence of Bose--Einstein
condensation (BEC) and the corresponding universality classes of the transition.

We consider the ideal Hamiltonian for $N$ particles
\begin{equation}
    H = \sum_{i=1}^N \left( \frac{\mathbf p_i^{2}}{2m} + V(\mathbf r_i) \right),\label{HN}
\end{equation}
where the external potential $V(\mathbf r)$ defines the geometry and the thermodynamic
limit via an extensive generalized volume ${\cal V}$. In this section, we assume that the potential, near its vanishing minimum, can be either a separable power law,
\begin{equation}
V(\mathbf{r}) \approx \sum_{i=1}^{d} V_{i}\, x_{i}^{s_{i}}, \label{Vgene}
\end{equation}
with \(V_{i} > 0\) and \(s_{i} > 0\), or an isotropic potential,
\begin{equation}
V(\mathbf{r}) \approx V_0\, r^{s} \>, \label{Viso}
\end{equation}
with $V_0 > 0$ and $r^2 = x_1^2 + \cdots + x_d^2$. Again, below we give specific examples of these generic forms.

\subsection{ Identifying the generalized volume}

For simplicity, let the ideal gas be confined by an external isotropic given by Eq. (\ref{Viso}). In the high-temperature limit the gas can be considered classical and the partition function for $N$ particles at a given temperature $T$ can be integrated to give, \cite{Sandoval2008}
\begin{eqnarray}
Z &=& \frac{1}{h^{dN}N!}\int d^{dN} r \int d^{dN} p \> e^{-H/k_BT} \nonumber \\
&=& \frac{1}{\lambda_T^{dN} N!} \left(\mathcal N_d (k_BT)^{d/s} V_0^{-d/s}\right)^N \>,
\end{eqnarray}
where $\mathcal N_d = \int d^d x e^{-x^s}$ is a number and $\lambda_T = h/\sqrt{2\pi m k_BT}$ the de Broglie wavelength. The free energy $F = -k_B T \ln Z$, for large $N$, is 
\begin{equation}
F = -Nk_BT \left( \ln\left[\frac{\mathcal N_d (k_BT)^{d/s}}{\lambda_T^d} \frac{V_0^{-d/s}}{N}\right] + 1\right) \>.
\end{equation}
One finds that $F/N$ is finite in the thermodynamic limit $F \to \infty$, $N \to \infty$, if and only if $V_0^{-d/s}/N$ also remains finite in such a limit. This requires that $V_0^{-d/s} \to \infty$. This allows us to define 
\begin{equation}
\mathcal V = V_0^{-d/s} 
\end{equation} as an extensive {\it volume}, generalizing $V$ for a box. Although the units need not those of a volume, this variable plays the role of a volume. In the same way, one defines a generalized {\it pressure} \cite{Sandoval2008,RomeroRochin2005} as $\mathcal P = -(\partial F/\partial {\mathcal V})_{N,T}$, such that ${\mathcal PV}$ is an energy. In particular, for the classical ideal gas,  ${\mathcal PV}= Nk_BT$, as expected.

\subsection{Semiclassical derivation of the low-energy DOS}

We begin with the semiclassical expression for the density of states
\begin{equation}
\rho(E)
= \frac{1}{(2\pi\hbar)^{d}}
\int d^{d}r \int d^{d}p\;
\delta\!\left(E - \frac{p^{2}}{2m} - V(\mathbf{r}) \right).
\end{equation}

 The integral over momentum space can be done, obtaining,
\begin{equation}
\rho(\epsilon)
= \frac{\Omega_{d}\,m}{(2\pi\hbar)^{d}}
\int_{V(\mathbf{r}) \le E} d^{d}r\,
\left[ 2m (E - V(\mathbf{r}))\right]^{\frac{d}{2}-1}.
\end{equation}
where \(\Omega_{d}\) is the solid angle in \(d\) dimensions.\\

Our interest is for small energies $\epsilon$ above the minimum of the potential that we assume has the separable power law form
given in Eq. (\ref{Vgene}). The density of states becomes,
\begin{equation}
\rho(\epsilon)
\approx \frac{\Omega_{d}\,m}{(2\pi\hbar)^{d}}
\int_{\cal D} d^{d}r\;
\left[
2m\left(\epsilon - \sum_{i=1}^{d} V_{i} x_{i}^{s_{i}}\right)
\right]^{\frac{d}{2}-1},
\end{equation}
where ${\cal D} = \{\mathbf{x}: \epsilon - \sum_{i=1}^{d}V_{i} x_{i}^{s_{i}} \ge 0\}$.\\

To make the low-energy scaling explicit, we rescale the spatial
coordinates as
\begin{equation}
y_{i} = \left( \frac{V_{i}}{\epsilon} \right)^{1/s_{i}} x_{i},
\qquad
d^{d}r = \epsilon^{\sum_{i} 1/s_{i}}
\prod_{i} V_{i}^{-1/s_{i}}\, d^{d}y
\end{equation}
yielding 
\begin{align}
\rho(\epsilon)
&\approx 
\Biggl[ \frac{\Omega_{d}\,m}{(2\pi\hbar)^{d}}
(2m)^{\frac{d}{2}-1}
\prod_{j=1}^{d} V_{j}^{-1/s_{j}} 
\int_{\mathcal{D}} d^{d}y\;
\left( 1 - \sum_{i=1}^{d} y_{i}^{s_{i}} \right)^{\frac{d}{2}-1}\Biggr] \epsilon^{\frac{d}{2}-1+\sum_{i}1/s_{i}}
\end{align}
where \(\mathcal{D} = \{ \mathbf{y} : \sum_{i} y_{i}^{s_{i}} \leq 1\}\). The term in brackets is a positive constant, independent of $\epsilon$, thus displaying the low energy scaling of the density of states, 
\begin{equation}
\rho(\epsilon) \sim \mathcal V \epsilon^{\,\frac{d}{2}-1 + \sum_{i} 1/s_{i}},
\end{equation}
where $\mathcal V\sim \prod_{j=1}^{d} V_{j}^{-1/s_{j}}.  $

For isotropic potentials $V(\mathbf{r}) \approx V_0\> r^s$, one gets the simpler, yet, clearer scaling,
\begin{equation}
\rho(\epsilon) \sim \mathcal V \epsilon^{\,\frac{d}{2}-1 + \frac{d}{s}},
\end{equation}
with a generalized volume  $\mathcal V\sim  V_0^{-d/s } $ as expected.

Therefore, the exponent $\sigma$ of the DOS, $\rho(\epsilon) \sim \mathcal V \> \epsilon^\sigma$, is either
\begin{eqnarray}
\sigma &=& \frac{d}{2}-1 + \sum_{i} 1/s_{i} \\
\sigma & = & \frac{d}{2}-1 + \frac{d}{s} \>.\label{sigmafin}
\end{eqnarray}

\section{ Density of States of representative confinement potentials}

Before discussing the details of specific trapping potentials, it is useful to summarize the relevant universality classes related to those cases. For convenience, Table~\ref{tab:geometries} lists the generalized volume, the DOS exponent~$\sigma$, and the corresponding universality class for each geometry.

\begin{table}[htbp]
\centering
\caption{Generalized volume ${\cal V}$, low-energy DOS exponent
$\sigma$, and resulting ideal-BEC universality class for
different confinement geometries. These classes are defined
in Table I of the main text.}
\label{tab:geometries}

\begingroup
\small
\renewcommand{\arraystretch}{1.3}
\setlength{\tabcolsep}{3pt}

\resizebox{\linewidth}{!}{%
\begin{tabular}{lccc}
\hline\hline
Geometry & Generalized volume ${\cal V}$
& DOS exponent $\sigma$ & Class \\
\hline

Box in $d$ dimensions
& $L^d$ & $\frac{d}{2}-1$
& \shortstack[l]{No BEC for 1D and 2D;\\ Class I for 3D} \\

Harmonic trap in $d$
& $\omega^{-d}$ & $d-1$
& \shortstack[l]{No BEC for 1D;\\ Class II for 2D;\\ Class III for 3D} \\

Quadrupolar trap in $d$
& $A^{-d}$ & $\frac{3d}{2}-1$
& \shortstack[l]{Class I for 1D;\\ Class III for 2D and 3D} \\

2D P\"oschl--Teller
& $L^2$ & $1$ & Class II \\

\shortstack[l]{3D power law\\
$ax^{s_x}+by^{s_y}$\\
$+cz^{s_z}$}
& $\begin{gathered}
a^{-1/s_x}\\
{}\times b^{-1/s_y}\\
{}\times c^{-1/s_z}
\end{gathered}$
& $\begin{gathered}
\frac{1}{s_x}+\frac{1}{s_y}\\
+\frac{1}{s_z}+\frac{1}{2}
\end{gathered}$
& \shortstack[l]{Class I, II, or III,\\ depending on $\sigma$} \\

\hline\hline
\end{tabular}%
}

\endgroup
\end{table}

\subsection{Rigid box in $d$ dimensions}

We first consider a hard-wall box of linear size $L$ in $d$ dimensions~\cite{LandauLifshitz},
\begin{equation}
V_\textrm{ext}(\vec r) =
\begin{cases}
0, & \vec r \in L^d, \\
\infty, & \vec r \notin L^d,
\end{cases}
\qquad
\textrm{(rigid box in $d$ dimensions)}.
\end{equation}
The natural thermodynamic variable is the spatial volume
\begin{equation}
    {\cal V} = L^d.
\end{equation}
The DOS in this case can be calculated and gives
\begin{equation}
    \rho(\epsilon) =
    L^d \frac{\tilde \Omega_d}{2(2\pi)^d}
    \left(\frac{2m}{\hbar^2}\right)^{d/2}\,
    \epsilon^{d/2-1},
    \qquad
    {\cal V}= L^d ,
\end{equation}
so that the DOS exponent is
\begin{equation}
    \sigma_\mathrm{box} = \frac{d}{2} - 1 \>,
\end{equation}
In particular, a 3D box yields $\sigma = 1/2$, whereas a 2D/1D box has $\sigma = 0, \,\text{and}\, \sigma =-1/2$ respectively and do not support finite-temperature ideal BEC. Here, the expression for $\sigma$, Eq. (\ref{sigmafin}), hold as
\begin{equation}
    \sigma=\frac{d}{2} - 1+\sum_i \frac{1}{s_i}=\frac{d}{2} - 1
\end{equation}
which, evidently, indicates that the exponents in Eqs. (\ref{Vgene}) and (\ref{Viso}) are $s_i \to \infty$ or $s \to \infty$.

\subsection{Isotropic harmonic oscillator (HO) in $d$ dimensions}

Next, we consider an isotropic harmonic trap~\cite{PitaevskiiStringari2003,PethickSmith2008},
\begin{equation}
 V_\textrm{ext}(\vec r)
 = \frac{1}{2} m \omega^2 (x_1^2 + \cdots + x_d^2),
 \qquad
 \textrm{(HO in $d$ dimensions)},
\end{equation}
with a single trap frequency $\omega$. The thermodynamic limit is reached using the generalized volume\cite{RomeroRochin2005}
\begin{equation}
    {\cal V} = \omega^{-d}
\end{equation}
as indicated above, see Eq. (\ref{Vgene}), without including the mass $m$ of particles. The DOS is
\begin{equation}
 \rho(\epsilon) =
 \frac{1}{\omega^d}
 \frac{2^{d-1} I_d \tilde \Omega^2_d}{(2\pi \hbar)^d}\,
 \epsilon^{d-1},
 \qquad
 {\cal V} = \frac{1}{\omega^d},
\end{equation}
with
\begin{equation}
I_d = \int_0^1 (1-x^2)^{d/2-1} x^{d-1} dx.
\end{equation}
Thus, the DOS exponent is
\begin{equation}
    \sigma_\mathrm{HO} = d - 1.
\end{equation}
In 3D, this gives $\sigma = 2$, placing the trapped ideal gas in the $\sigma > 1$ universality class
where the correlation length diverges with $\nu = 1/2$ but the compressibility remains finite. For 2D we have BEC in the Class II universality class. \cite{MoralesAmador2024} Finally, no ideal BEC takes place for 1D.
Here, we notice that we can write 
\begin{equation}
    \sigma_\mathrm{HO}=\frac{d}{2} - 1+\sum_{i=1}^d \frac{1}{s_i}= d - 1
\end{equation}
with $s_i=2$. Which recovers precisely the expression given in Eq. (\ref{sigmafin}).

\subsection{ Quadrupolar (linear-radial) confinement in $d$ dimensions}

We now consider a radially linear (“quadrupolar”) confinement~\cite{Sandoval2008},
\begin{equation}
 V_\textrm{ext}(\vec r)
 = A (x_1^2 + \cdots + x_d^2)^{1/2},
 \qquad
 \textrm{(quadrupolar in $d$ dimensions)},
\end{equation}
with $A > 0$ setting the trap strength. The generalized volume in this case, see Eq. (\ref{Vgene}), is
\begin{equation}
    {\cal V} = \frac{1}{A^d}.
\end{equation}
In this case, the DOS is given by
\begin{equation}
 \rho(\epsilon) =
 \frac{1}{A^d}
 \frac{J_d \tilde \Omega^2_d}{2(2\pi)^d}
 \left(\frac{2m}{\hbar^2}\right)^{d/2}
 \epsilon^{3d/2 - 1},
 \qquad
 {\cal V} = \frac{1}{A^d},
\end{equation}
where
\begin{equation}
J_d = \int_0^1 (1-x)^{d/2-1} x^{d-1} dx.
\end{equation}
The low-energy DOS exponent is therefore
\begin{equation}
    \sigma_\mathrm{quad} = \frac{3d}{2} - 1.
\end{equation}
Such strongly confining traps provide a way to realize large $\sigma$, and hence to explore regimes
deep in the $\sigma>1$ class for 2D and 3D where only the correlation length exhibits universal critical behavior. For 1D, BEC falls in the Class I universality class.

In this case, 
\begin{equation}
    \sigma_\mathrm{quad} =\frac{d}{2} - 1+\sum_{i=1}^d \frac{1}{s_i}= \frac{3d}{2} - 1
\end{equation}
with $s_i=1,$ in agreement with our analysis. 

\subsection{Two-dimensional Pöschl--Teller (PT) potential}

A key example with a nontrivial but \emph{analytic} DOS at threshold is the two-dimensional
Pöschl--Teller (PT) potential,\cite{Kleebank:2025}
\begin{equation}
 V_\textrm{ext}(\vec r)
 = \frac{\hbar^2 \lambda(\lambda-1)}{2m}
   \left(\frac{\pi}{L}\right)^2
   \left[
      \frac{1}{\cos^2 \left(\frac{\pi x_1}{L}\right)}
    + \frac{1}{\cos^2 \left(\frac{\pi x_2}{L}\right)}
   \right],
 \qquad
 \textrm{(PT in $d = 2$)} \>,
\end{equation}
where $\lambda > 1$ is a parameter and $L$ the asymptotic size of the potential.
The natural generalized volume turns out to be area,
\begin{equation}
    {\cal V} = L^2 \>,
\end{equation}
which can be seen to be consistent with Eq. (\ref{sigmafin}) by expanding the potential near its minimum and by identifying  \cite{Kleebank:2025},
\begin{equation}
    V_0 = \frac{\hbar^2}{2m}\left(\frac{\lambda}{\pi L}\right)^2,
\end{equation}
the single particle ground state, as an {\it intensive} variable~\cite{Kleebank:2025}. 
The DOS for this potential reads
\begin{equation}
 \rho(\epsilon) =
 L^2  \frac{m}{\pi^2 \hbar^2}
 \left[
   \arctan\!\left( \sqrt{\frac{\epsilon + V_0}{V_0}} \right)
   - \arctan\!\left( \sqrt{\frac{V_0}{\epsilon + V_0}} \right)
 \right],
 \qquad
 {\cal V} = L^2 .
\end{equation}
Importantly, the DOS is analytic at $\epsilon = 0$, and its leading low-energy behavior is
\begin{equation}
 \rho(\epsilon) \approx
 L^2 \frac{m}{2\pi^2 \hbar^2}\,
 \frac{\epsilon}{V_0},
 \qquad (\epsilon \ll V_0),
\end{equation}
from which we identify
\begin{equation}
    \sigma_\mathrm{PT,2D} = 1.
\end{equation}
Thus, the 2D PT potential realizes the marginal case $\sigma = 1$, which belongs to the Class II
universality class characterized by logarithmic corrections to scaling. While PT potentials closely resemble a box potential, their small softening allows to support ideal BEC in 2D~\cite{Kleebank:2025}. In this case, 
\begin{equation}
    \sigma_\mathrm{PT,2D} = \frac{d}{2}-1+\sum_{i=1}^2\frac{1}{s_i}=1
\end{equation}
with $d=2$ and $s_i=2.$ 

\subsection{ Generic 3D power-law confinement}

Finally, we consider a generic separable power-law trap in three dimensions~\cite{de1950bose,Bagnato1987},
\begin{equation}
 V_\textrm{ext}(\vec r)
 = a\, x^{s_x} + b\, y^{s_y} + c\, z^{s_z},
 \qquad
 \textrm{(generic power-law in $d = 3$)},
\end{equation}
with $a,b,c>0$ and exponents $s_x,s_y,s_z>0$. The generalized volume is
\begin{equation}
    {\cal V} = \frac{1}{a^{1/s_x} b^{1/s_y} c^{1/s_z}}.
\end{equation}
The DOS can be written as
\begin{equation}
 \rho(\epsilon) =
 \frac{1}{a^{1/s_x}\, b^{1/s_y}\, c^{1/s_z}}
 \frac{K_3}{(2\pi)^2}
 \left(\frac{2m}{\hbar^2}\right)^{3/2}
 \epsilon^{1/s_x + 1/s_y + 1/s_z + 1/2},
 \qquad
 {\cal V} = \frac{1}{a^{1/s_x}\, b^{1/s_y}\, c^{1/s_z}},
\end{equation}
with
\begin{equation}
 K_3 =
 \int_{-1}^1 (1-x^{s_x})^{1/2+1/s_y+1/s_z} dx
 \int_{-1}^1 (1-y^{s_y})^{1/2+1/s_z} dy
 \int_{-1}^1 (1-z^{s_z})^{1/2} dz.
\end{equation}
We thus identify the DOS exponent as
\begin{equation}
    \sigma_\mathrm{pow,3D} =\sum_{i=x,y,z} \frac{1}{s_i} + \frac{1}{2}=\frac{d}{2}-1+\sum_i \frac{1}{s_i},
\end{equation}
which is precisely formula Eq. (\ref{sigmafin}) for $d=3.$  By tuning the exponents $s_i$, this class of traps allows one to engineer essentially arbitrary
values of $\sigma > 0$ and hence to access different universality classes of ideal BEC in a controlled
way.


\section{ Two-particle Green's function, LDA, and trapped correlations}
In this section, we study the decay of the density-density correlations within the local density approximation (LDA) for trapped systems. We show that the DOS governs the behaviour of long-range decay of the correlations.

\subsection{ Single-particle Green's function}
Our starting point is the single-particle Green's function at equal time. We first assume an ideal Bose gas in $d$ dimensions confined in a homogeneous trap,
\begin{equation}
G_{1,\textrm{hom}}(\mathbf r) = \frac{1}{(2\pi)^d}\int d^d k \frac{e^{i \mathbf k \cdot \mathbf r}}{e^{(\epsilon_k - \mu)/k_BT}-1} \>,
\end{equation}
where $r = |\mathbf x - \mathbf x'|$ is the relative separation of the correlated points and $\epsilon_k = \hbar^2 k^2/2m$. It is convenient to introduce the  correlation length
\begin{equation}
 \xi^{-1} = \sqrt{\frac{2m|\mu|}{\hbar^2}},
\end{equation}
which is the dominant length in the vicinity of the onset of BEC, namely, for $\mu \to 0^-$. In this limit, the homogeneous
single–particle Green’s function can be approximated,
\begin{equation}
G_{1,\textrm{hom}}(\mathbf r) \approx \frac{1}{(2\pi)^d}k_BT\int d^d k \frac{e^{i \mathbf k \cdot \mathbf r}}{\epsilon_k - \mu} \>,
\end{equation}
which can be exactly integrated, giving
\begin{equation}
G_{1,\mathrm{hom}}(r) \sim
\left( \frac{1}{\xi r} \right)^{\frac{d}{2}-1}
K_{\frac{d}{2}-1}(r/\xi),
\label{eq:G1_hom}
\end{equation}
where 
$K_\nu$ is the modified Bessel function of the second kind. Up to an overall temperature dependent prefactor, not relevant for critical properties,
Eq.~\eqref{eq:G1_hom} describes the spatial decay of single–particle
correlations in a homogeneous ideal Bose gas near BEC.

Here, we should note that for $r/\xi  \ll 1$
we find that 
\begin{equation}
G_{1,\mathrm{hom}}(r) \sim \frac{1}{r^{d-2+\eta}}
\end{equation}
with $\eta$ the so-called anomalous dimension of the field which is stricly zero $\eta=0$ for non-interacting homogenoues BEC, of course, as expected.

In what follows, we will use Eq.~\eqref{eq:G1_hom} as the local building block
to construct the two–particle Green’s function in a trapped gas within the
LDA.



\subsection{ Density--density correlations}
As discussed in the main text, the \emph{two–particle} (density–density) Green’s function is the relevant object for thermodynamic response functions. We will employ a local-density approximation, for this purpose it is useful to change to the center–of–mass
and relative coordinates,
\begin{equation}
\mathbf r = \mathbf x - \mathbf x', \qquad
\mathbf R = \frac{\mathbf x + \mathbf x'}{2},
\end{equation}
so that $G(\mathbf x,\mathbf x') = G(\mathbf r,\mathbf R)$.
For an ideal gas, Wick’s theorem implies that the local two–particle
correlator can be written as

\begin{equation}
G_2(\mathbf r,\mathbf R) = \rho(\mathbf R) \delta^d(\mathbf r) + \bigl|G_1(\mathbf r,\mathbf R)\bigr|^2 \>,
\end{equation}
where $\rho(\mathbf R)$ is the local density. This correlation function is exactly related to the isothermal compressibility by,
\begin{equation}
\kappa_T = \frac{\mathcal V}{N^2 k_BT} \int d^d r\int d^d R\, G_2(\mathbf r,\mathbf R) \>.
\label{eq:kappa_fluc}
\end{equation}
Near BEC, $\kappa_T$ takes large values and the relevant contribution of the correlation are for large values of $r$, the distance between the two correlated spatial points. Thus, the relevant quantity to study is the center-of-mass integrated correlation function, that is,

\begin{equation}
\mathcal G(\mathbf r) = \int d^d R\, G_2(\mathbf r,\mathbf R)
= \int d^d R\, \bigl|G_1(\mathbf r,\mathbf R)\bigr|^2 \>.
\label{eq:g2_def}
\end{equation}

\subsection{Local density approximation for the trapped gas}

In a trapped system translational invariance is broken and
$G_1(\mathbf x,\mathbf x') \neq G_1(\mathbf x-\mathbf x')$. Within the
local density approximation, one assumes that around each point $\mathbf R$
the system behaves as a homogeneous gas characterized by a \emph{local}
chemical potential,
\begin{equation}
\mu_{\mathrm{LDA}}(\mathbf R) = \mu - V(\mathbf R),
\label{eq:mu_LDA}
\end{equation}
where $V(\mathbf R)$ is the trapping potential. The corresponding local
inverse coherence length is
\begin{equation}
\xi^{-1}(\mathbf R)
= \sqrt{\frac{2m\,|\mu_{\mathrm{LDA}}(\mathbf R)|}{\hbar^2}} = \sqrt{\xi^{-2}+ \frac{2m V(\mathbf R)}{\hbar^2}}\>.
\end{equation}

Within LDA, the inhomogeneous single–particle Green’s function is then
approximated by the homogeneous one evaluated at the local scale $\xi(\mathbf R)$,
\begin{equation}
G_{1}(\mathbf r,\mathbf R) \simeq G_{1,\mathrm{hom}}\!\left(\mathbf r; \xi(\mathbf R)\right),
\end{equation}
with $G_{1,\mathrm{hom}}$ given by Eq.~\eqref{eq:G1_hom}. Consequently, the
two–particle Green’s function in the trapped system is
\begin{equation}
G_{2}(\mathbf r,\mathbf R) =\bigl|G_{1,\mathrm{hom}}(\mathbf r;\xi (\mathbf R))\bigr|^2.
\end{equation}
Our task is to analyze the asymptotic behavior of the averaged correlator
$\mathcal G(\mathbf r)$ in Eq.~\eqref{eq:g2_def} for a power–law trap
$V(\mathbf R)=V_0 R^s$. Now, we study two different regimes, the large-distance regime $r/\xi\gg 1$, for large but finite $\xi$, and at criticality $r/\xi\ll 1$.\\

\subsection{ Asymptotics for $r/\xi \gg 1$}

We first focus on the large–distance regime $r/\xi \gg 1$. In this limit, the modified Bessel function admits the asymptotic expansion
\begin{equation}
K_{\nu}(x)\sim \sqrt{\frac{\pi}{2x}}\,e^{-x}
\qquad (x\gg 1),
\end{equation}
so that, using Eq.~\eqref{eq:G1_hom}, the leading form of the \emph{two–particle}
Green’s function is
\begin{equation}
G_{2}(\mathbf r,\mathbf R) \sim 
\left(\frac{1}{\xi(\mathbf R) r}\right)^{d-2}
\frac{\xi(\mathbf R)}{r}\,e^{-2r/\xi(\mathbf R)}.
\label{eq:G2_asymp}
\end{equation}

For the spherically symmetric power–law trap $
V(R)=V_0 R^s$,
we can expand the local correlation length around its central value $\xi$, allowed by the thermodynamic limit $V_0 \to 0$,
\begin{equation}
\xi^{-1}(\mathbf R)\approx \xi^{-1} + \frac{m V_0}{\hbar^2}\xi R^s \>.
\end{equation}

Substituting into Eq. (\ref{eq:G2_asymp}) and integrating over the center of mass coordinate $\mathbf R$, one obtains, see Eq. (\ref{eq:g2_def}),
\begin{equation}
\mathcal G(r)\sim
\mathcal V\;
\frac{\xi^{-(d-3+d/s)}}{r^{\,d-1+d/s}
}\,e^{-2r/\xi} \>,
\label{eq:g2_xi_large}
\end{equation}
which is Eq. (22) of the main text. This expression gives the very relevant result that $\xi/2$ is the correlation length near BEC for any confining potential, a divergent length $\xi \sim t^{-\nu_T}$ as $\mu \to 0^-$. In the main text we have found the exponent $\nu_T$ in terms of the exponent $\sigma$ of the DOS. Note that $\mathcal G(r)$ is proportional to the generalized volume $\mathcal V = V_0^{-d/s}$; this is an important observation because its presence renders the compressibility as an intensive quantity, see Eq. (\ref{eq:kappa_fluc}).

\subsection{ Asymptotics for $ r/\xi \ll 1$}

We now consider the opposite limit
\begin{equation}
 r/\xi \ll 1 .
\end{equation}
This is the situation where the system sits at criticality $\mu = 0$, then the local coherence length is 
\begin{equation}
\xi^{-1}(\mathbf R) = \left(\frac{2m V_0 R^s}{\hbar^2}\right)^{1/2} \>.
\end{equation}
Substitution into the LDA correlation function $G_{2}(\mathbf r, \mathbf R)$, gives
\begin{equation}
G_{2}(\mathbf r,\mathbf R)
\sim
\left(\frac{V_0 R^{s/2}}{r}\right)^{d-2}
K_{\frac{d}{2}-1}^{\,2}\!\bigl(\sqrt{\frac{2mV_0}{\hbar^2}} R^{s/2} r\bigr).
\end{equation}
After integrating out the center-of-mass coordinate one gets,
\begin{equation}
\mathcal G(r)
\sim
\frac{\mathcal V}{r^{\,2d-4+\frac{2d}{s}}}
\int_0^\infty d z\;
z^{\frac{2d}{s}-3+d}
K_{\frac{d}{2}-1}^{\,2}(z),
\end{equation}
where the remaining integral over $z$ is finite and dimensionless. From this expression one can read off the $\eta_T$ exponent by appealing to Fisher form
\begin{equation}
\mathcal G(r)\sim \frac{1}{r^{\,d-2+\eta_T}},
\end{equation}
thus identifying
\begin{equation}
\eta_T = d-2+\frac{2d}{s}.
\end{equation}
Equivalently,
\begin{equation}
\eta_T = 2\sigma.
\end{equation}
\end{widetext}

\bibliographystyle{apsrev4-2}
\bibliography{references}

\end{document}